\documentclass[print,twocolumn,showpacs,preprintnumbers,superscriptaddress,amsmath,amssymb]{revtex4}
\usepackage{amsmath}
\usepackage{color,soul}
\usepackage{graphicx}
\usepackage{dcolumn}% Align table columns on decimal point
\usepackage{bm}
\usepackage[dvipdfm,pdfstartview=FitH]{hyperref}

\begin{document}

\title{Many-qubit network employing cavity QED in a decoherence-free subspace}

\author{Hua Wei}
\email{huawei.hw@gmail.com}

\affiliation{State Key Laboratory of Magnetic Resonance and Atomic and Molecular Physics,
Wuhan Institute of Physics and Mathematics, Chinese Academy of Sciences,
Wuhan, 430071, China}
\affiliation{Graduate School of the Chinese Academy of Sciences, Beijing, 100049, China}

\author{Wan Li Yang}
\affiliation{State Key Laboratory of Magnetic Resonance and Atomic
and Molecular Physics, Wuhan Institute of Physics and Mathematics,
Chinese Academy of Sciences, Wuhan, 430071, China}
\affiliation{Graduate School of the Chinese Academy of Sciences,
Beijing, 100049, China}

\author{Zhi Jiao Deng}
\affiliation{State Key Laboratory of Magnetic Resonance and Atomic and Molecular Physics,
Wuhan Institute of Physics and Mathematics, Chinese Academy of Sciences,
Wuhan, 430071, China}
\affiliation{Graduate School of the Chinese Academy of Sciences, Beijing, 100049, China}

\author{Mang Feng}
\email{mangfeng@wipm.ac.cn}

\affiliation{State Key Laboratory of Magnetic Resonance and Atomic
and Molecular Physics, Wuhan Institute of Physics and Mathematics,
Chinese Academy of Sciences, Wuhan, 430071, China}

\begin{abstract}
We propose a many-qubit network with cavity QED by encoding qubits
in decoherence-free subspace, based on which we can implement
many-logic-qubit conditional gates by means of cavity assisted
interaction with single-photon pulses. Our scheme could not only
resist collective dephasing errors, but also much reduce the
implementational steps compared to conventional methods doing the
same job, and we can also complete communications between two
arbitrary nodes. We show the details by implementing a
three-logic-qubit Toffoli gate, and explore the experimental
feasibility and challenge based on currently achievable cavity QED
technologies.
\end{abstract}

\maketitle

In quantum information science, one-qubit rotations and two-qubit
conditional operations could constitute universal quantum computing
(QC) \cite{de}. Although these basic operations have been achieved
experimentally in various systems, to have a large-scale QC
efficiently with high-fidelity, we are still exploring direct
accomplishment of many-qubit conditional gates for simplifying the
operational steps and decreasing the implementing time.

Cavity QED system has been considered as a wonderful device to
realize the quantum information processing. The confined atoms in
cavity QED system are not only suited for storing qubits with
long-lived atomic internal states, but also usable for repeaters in
quantum networks \cite{transfer,cpf}. We have noticed recent
proposals for multi-atom quantum gates carried out in a single
cavity \cite{chen}and in a network of non-local cavities
\cite{deng}, which provide hopeful ways for scalable QC with cavity
QED.

However, decoherence due to inevitable interaction with environment
always damages quantum gating. In this Brief Report, we will focus
on avoiding collective dephasing errors by decoherence-free subspace
(DFS) \cite{dfs0,dfs1,dfs2}, in which some unpredictable collective
dephasing due to, for example, ambient magnetic fluctuations, would
be kept away from our encoding qubits. To get this benefit, we have
to pay a price with two physical qubits encoding one logic qubit.
Specifically, in our present case, two atoms in one cavity encode a
single logic qubit \cite{dfs2}, with the form $|\widetilde{0}\rangle
\equiv |1\rangle _{1}|0\rangle _{2}=|10\rangle $ and
$|\widetilde{1}\rangle \equiv |0\rangle _{1}|1\rangle
_{2}=|01\rangle $.

With this encoding, we will propose a nonlocal many-qubit gating in
DFS in a quantum network constituted by cavities, based on
cavity-assisted interaction by single-photon interference. To make
our description clarified, we will take the three-logic-qubit
Toffoli gate as an example, and our idea is directly applicable to
the case of arbitrary numbers of logic qubits. The favorable
features of our scheme include a big reduction of the implementational
steps compared to conventional methods by the network of one- and
two-qubit quantum gates. Besides, the Hadamard operation of the
single-logic-qubit and multi-logic-qubit conditional operations
could coexist in our design but work independently due to our
elaborate control of the path of the single-photons, which helps for
compatibility of the quantum network. Furthermore, our network
allows quantum communications between arbitrary two cavity nodes.

\begin{figure}[tbp]
\includegraphics[width=6.0cm]{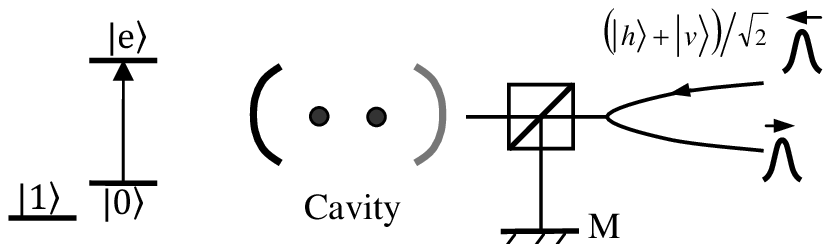}
\caption{Left: Level configuration of an atom in our scheme, where the two
ground states $|0\rangle$ and $|1\rangle$ with physical qubits encoded
are of large energy splitting, and $|e\rangle$ is the excited
state. Right: Two identical atoms are well located in a high-finesse
single-mode cavity to implement a two-atom $U_{\text{CPF}}$ by
cavity-assisted polarized-photon scattering. The polarizing beam
splitter (PBS) transmits (reflects) the $h$ ($v$) component of an
input single-photon pulse. \label{1}}
\end{figure}

The fundamental cavity-assisted photon scattering to realize a
conditional phase flip (CPF) between two atoms in a cavity has been
reiterated in \cite{cpf,wei,deng}. As shown in Fig.~\ref{1}, the
cavity mode, the input photon pulse and the transition between the
levels $|0\rangle$ and $|e\rangle$ of the atom are resonantly
coupled, while level $|1\rangle$ is decoupled owing to the large
detuning. Because the cavity mode is $h$ polarized, it only
interacts with the $h$ component of an input photon. The atom in
$|0\rangle$ will shift the cavity mode so that the photon pulse will
leave the cavity with nothing changed. While the atom in $|1\rangle$
leads to the $\pi$ phase added when the photon left because of the
resonance between the photon and the cavity mode. The key point of the
CPF is that an input single-photon pulse with $h$ polarization could
interact with the cavity mode if and only if the two atoms are in state
$|1\rangle_{1}|1\rangle _{2}$. As a result, the single-photon pulse
moving in and then out of the cavity yields a two-atom CPF operation
$U_{\text{CPF}}=\exp (i\pi|1\rangle_{11}\langle1|\otimes |1\rangle_{22}\langle 1|)$.

\begin{figure*}[tbp]
\includegraphics[width=\textwidth]{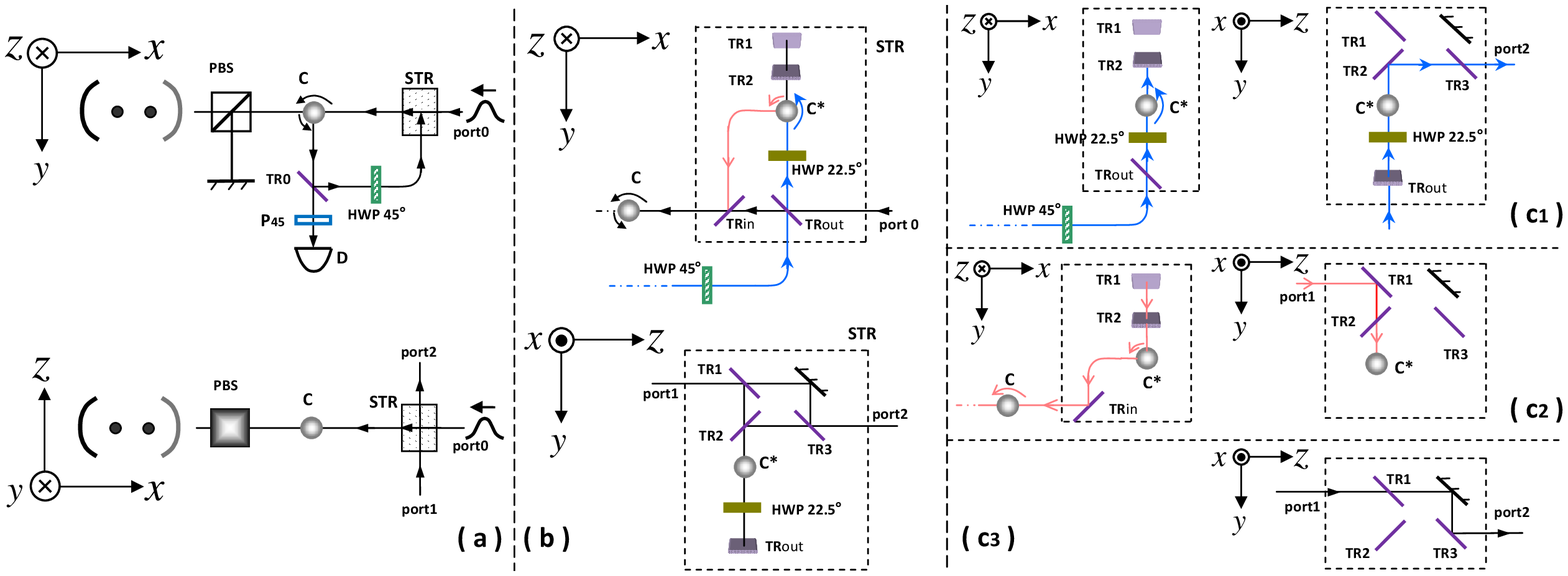}
\caption{(Color online) Schematic setup of the basic node unit of
our design from different views. (a) Setup viewed along $z$-axis and
viewed on the $x-z$ plane; (b) The configuration of STR, which
consists of several optical elements, such as TR (marking in purple)
, circulator and haft-wave plate. STR can control the path of the
photon to different planes (see text for detail). The single-photon
pulse can enter the cavity node directly from the port 0 (along the
black arrow); (c) The photon can be guided in the cavity node from
the port 1 (along the red thin arrow), and be guided out of the
cavity node from the port 2 (along the blue fat arrow). Furthermore,
the photon can cross STR directly between port 1 and port 2.
HWP$45^\circ$ rotates the photon polarization as
$|h\rangle\leftrightarrow|v\rangle$, HWP$22.5^\circ$ perform
Hadamard operation on the photon state. D is a detector, C and C*
are circulators, P$_{45}$ is a $45^\circ$ polarizer. TR can be
controlled exactly as needed to transmit or to reflect a photon.}
\label{2}
\end{figure*}

With the cavity-assisted photon scattering, both the
single-logic-qubit Hadamard gate $\widetilde{H}$ and the
two-logic-qubit conditional phase gate $\widetilde{U}_{\text{CP}2}$
had been realized between two neighboring nodes in a cavity-based
system \cite{wei}, where $\widetilde{H}$ depends on twice
interactions of the photon with the cavity and
$\widetilde{U}_{\text{CP}2}$ needs only one interaction with each
cavity. While to extend our idea to be more than two cavities, we
need to design a more smart device. Fig.~\ref{2} demonstrates from
different views the basic unit of such a design, where the half-wave
plate 45$^\circ$ (HWP45$^\circ$), with its axis at $45^\circ$ to the
horizontal direction, rotates the photon polarization as
$|h\rangle\leftrightarrow|v\rangle$. HWP$22.5^\circ$, at an angle of
$22.5^\circ$ to the horizontal direction, performs a Hadamard gate
on the photon polarization states, i.e.,
$|h\rangle\leftrightarrow(1/\sqrt{2} )(|h\rangle+|v\rangle)$,
$|v\rangle\leftrightarrow(1/\sqrt{2} )(|h\rangle-|v\rangle)$. C and
C* are circulators. P$_{45}$ is a $45^\circ$ polarizer projecting
the polarization $(|h\rangle+|v\rangle)/\sqrt{2}$. The different
graphic denotations of PBS and TR are due to different views.

TR, marked in purple, including TR0, TR1, TR2, TR3, TRin and TRout
in Fig.~\ref{2}, are optical devices which can be controlled exactly
as needed to transmit or reflect a photon with a very fast switching
time. All the TR devices are identical, but labeled to be TR${\it
m}$ $({\it m}=0, 1, 2, \cdots)$ for convenience of our description.
STR, a special device which controls the path of the photon to
different planes, has three ports allowing the single-photon pulses
to communicate with the cavity node. The arrows in different color
in Fig.~\ref{2} show the different photon paths regarding the three
ports, respectively. The single-photon pulse can enter the cavity
node directly from the port 0 along the black arrow depending on the
transmitting states of TRin and TRout, as shown in the top plot of
Fig.~\ref{2}(b). Alternatively, STR allows the photon input from the
port 1 following the red thin arrow (Fig.~\ref{2}(c2)). After its
interacting with the atoms in the cavity, the single-photon pulse
should move along the blue fat arrow going to TR2 and then be
reflected out of the node from the port 2 (Fig.~\ref{2}(c1)).
Moreover, if it just passes by the cavity node, the single-photon
pulse will pass through STR from the port 1 directly to the port 2
with the help of TR1 and TR3, as shown in Fig.~\ref{2}(c3).

Within the cavity node, the movement of the single-photon pulse also
depends on the state of TR0. When TR0 is switched to a transmitting
state, the photon will pass through TR0 and P$_{45}$ and then be
measured by the detector directly. This is part of the way for
$\widetilde{H}$ \cite {wei}. To carry out conditional gates, we
require TR0 to keep in a reflecting state, which makes the photon go
out of the cavity node along the blue fat arrow from the port 2. To
implement our scheme, we have to switch the relevant TR into the
corresponding transmitting or reflecting state according the photon
paths in Fig.~\ref{2}.

\begin{figure}[tbp]
\includegraphics[width=8.5cm]{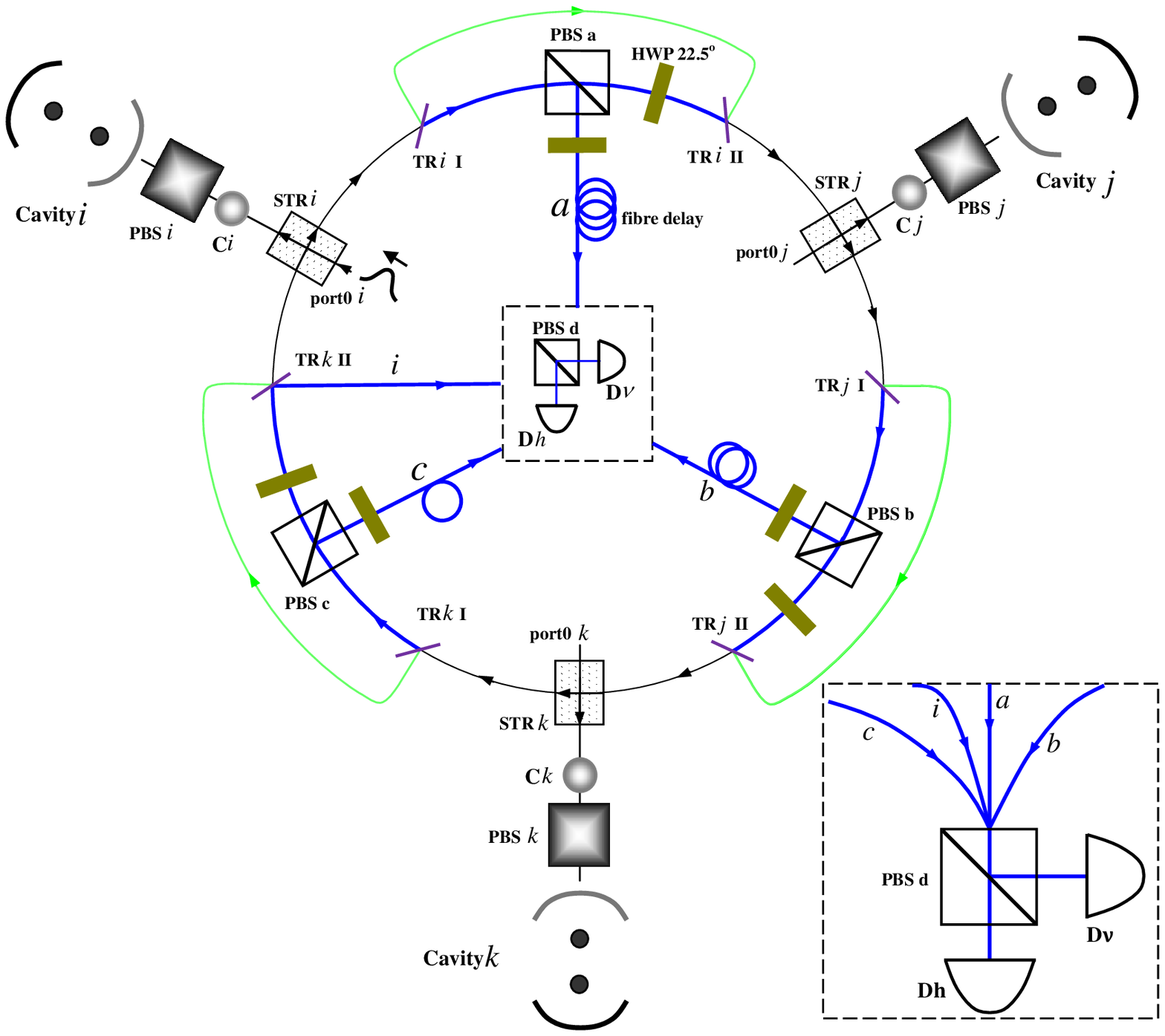}
\caption{(Color online) Schematic setup for implementation of
three-logic-qubit conditional phase gate
$\widetilde{U}_{\text{CP}3}$ by single-photon interference in
clock-wise direction. The photon will take four possible paths $a$,
$b$, $c$ and $i$ to reach PBS$d$. The single-logic-qubit Hadamard
operation and the two- and multi-logic-qubit conditional operation
are coexisting in our scheme but could work independently by
controlling the path of the single-photons. The thick blue circle is
the necessary path for the $\widetilde{U}_{\text{CP}3}$ operation,
and the light green path for the $\widetilde{U}_{\text{CP}2}$
between two logic qubits. Each dark-green half-wave plate stands for
a HWP $22.5^\circ$. The inset is for the detection made at the
center of the network.} \label{3}
\end{figure}

Based on above basic unit, we construct a circular many-qubit cavity
network in DFS in Fig.~\ref{3}. To make our description clarified
and simple, without loss of generality, we only demonstrate a
three-logic-qubit Toffoli gate
$\widetilde{T}_{\text{offoli}}^{ij,k}=\widetilde{H}_k\otimes
\widetilde{U}_{\text{CP}3}\otimes\widetilde{H}_k$ as an example,
where $\widetilde{H}_k$ and $\widetilde{U}_{\text{CP}3}$ are
Hadamard gate on the $k$th logic-qubit and three-logic-qubit
conditional phase gate, respectively. For the many-qubit conditional
gate, the key point is not the order of the operations on individual
nodes, but the phase flip when all target logic-qubits are in states
$|\widetilde{1}\rangle$. So as an example, we give in Fig.~\ref{3}
an implementation of $\widetilde{U}_{\text{CP}3}$ by a clock-wise
operations based on the single-photon interference.

The operation of $\widetilde{U}_{\text{CP}3}$ as shown in
Fig.~\ref{3} will follow the thick blue optical path on the circle.
It can be performed by following steps. A single-photon pulse in
state $(1/\sqrt{2})(|h\rangle+|v\rangle)$ is imported from the port
0$i$, i.e., the port 0 belonging to the $i$th node. As mentioned
previously in Fig.~\ref{2}, the input photon will enter the node$i$
directly through STR$i$, and reach the cavity$i$ through C and PBS.
To have the operation $U_{\text{CPF}}$, we perform a $\sigma_x$
operation on the atom 2 in the cavity$i$ (denoted as
$\sigma_{x,i}^2$) so that $U_{\text{CPF}}$ could happen when the
single-photon pulse moves in and then out of the cavity. Then we
carry out another $\sigma_x$ operation to restore the state of the
atom 2. Subsequently, the photon moves back to C again, turned by C
and then reflected by TR0 to HWP$45^\circ$. Finally, the
single-photon pulse moves out of the node$i$ from the port 2$i$ of
STR$i$. Let us turn to Fig.~\ref{3} again. After the single-photon
pulse moves out of the node$i$, it clock-wisely arrives at PBS$a$
through TR$i$I along the thick blue line. Whether the photon pulse
will pass through PBS$a$ to node$j$ or be reflected along the
path$a$ to PBS$d$ depends on the state of the logic-qubit insides
the node$i$ (i.e., cavity$i$). Specifically, for the logic-qubit in
cavity$i$ in state $| \widetilde{0}\rangle_i$ or
$|\widetilde{1}\rangle_i$, the above photon path in the basic unit
leads to
\begin{equation}
\begin{split}
&\frac{|h\rangle+|v\rangle}{\sqrt{2}}|\widetilde{0}\rangle_i
\xrightarrow{\sigma_{x,i}^2U_{\text{CFP}}\sigma_{x,i}^2}\frac{
-|h\rangle+|v\rangle}{\sqrt{2}}|\widetilde{0}\rangle_i\\
&\xrightarrow{\text{HWP}45^\circ}\frac{|h\rangle-|v\rangle}{\sqrt{2
}}|\widetilde{0}\rangle_i\xrightarrow{\text{HWP}22.5^\circ}|v\rangle|\widetilde{0}
\rangle_i.
\end{split}
\label{eq1}
\end{equation}
\begin{equation}
\begin{split}
&\frac{|h\rangle+|v\rangle}{\sqrt{2}}|\widetilde{1}\rangle_i
\xrightarrow{\sigma_{x,i}^2U_{\text{CFP}}\sigma_{x,i}^2}\frac{
|h\rangle+|v\rangle}{\sqrt{2}}|\widetilde{1}\rangle_i\\
&\xrightarrow{\text{HWP}45^\circ}\frac{|h\rangle+|v\rangle}{\sqrt{2
}}|\widetilde{1}\rangle_i\xrightarrow{\text{HWP}22.5^\circ}|h\rangle|\widetilde{1}
\rangle_i.
\end{split}
\label{eq2}
\end{equation}

So if the logic-qubit in cavity is in state $|\widetilde{0}\rangle$,
the single-photon pulse will be reflected by PBS$a$ to go along the
path$a$, whereas the photon will go through PBS$a$ and
HWP$22.5^\circ$ to the node$j$ if the atoms are in
$|\widetilde{1}\rangle$. For the latter case, the $h$-polarized
photon will go through a HWP$22.5^\circ$ before entering the cavity
node$j$, and thereby becomes in superposition
$(1/\sqrt{2})(|h\rangle+|v\rangle)$ again. The interaction between
the photon pulse and the atoms in cavity$j$ is similar to in the
cavity$i$, and the subsequent process to the cavity$k$ is similar to
that to the cavity$j$. In our three-logic-qubit case, the whole
optical path ends at reaching the detector D$h$ or D$v$ along the
path$i$. The click of D$v$ yields a minus sign in front of the state
$|\widetilde{1}\rangle_i|\widetilde{1}\rangle_j|\widetilde{
1}\rangle_k$, and the click of D$h$ means nothing changed.
Straightforward deduction could show that, for three logic-qubits
initially in an arbitrary state $\beta_1|\widetilde{0}\rangle_i|
\widetilde{0}\rangle_j|\widetilde{0}\rangle_k+ \beta_2|\widetilde{0}
\rangle_i|\widetilde{0}\rangle_j|\widetilde{1}\rangle_k+
\beta_3|\widetilde{0
}\rangle_i|\widetilde{1}\rangle_j|\widetilde{0}\rangle_k+
\beta_4|\widetilde{
0}\rangle_i|\widetilde{1}\rangle_j|\widetilde{1}\rangle_k+ \beta_5|
\widetilde{1}\rangle_i|\widetilde{0}\rangle_j|\widetilde{0}\rangle_k+
\beta_6|\widetilde{1}\rangle_i|\widetilde{0}\rangle_j|\widetilde{1}
\rangle_k+
\beta_7|\widetilde{1}\rangle_i|\widetilde{1}\rangle_j|\widetilde{0
}\rangle_k+
\beta_8|\widetilde{1}\rangle_i|\widetilde{1}\rangle_j|\widetilde{
1}\rangle_k$, the above process yields
\begin{equation}
\begin{split}
&\frac{|h\rangle}{\sqrt{2}}\otimes
(\beta_1|\widetilde{0}\rangle_i|\widetilde{0}\rangle_j|\widetilde{0}\rangle_k+
\beta_2|\widetilde{0}\rangle_i|\widetilde{0}\rangle_j|\widetilde{1}\rangle_k+
\beta_3|\widetilde{0}\rangle_i|\widetilde{1}\rangle_j|\widetilde{0}\rangle_k\\
&+\beta_4|\widetilde{0}\rangle_i|\widetilde{1}\rangle_j|\widetilde{1}\rangle_k
+\beta_5|\widetilde{1}\rangle_i|\widetilde{0}\rangle_j|\widetilde{0}\rangle_k+
\beta_6|\widetilde{1}\rangle_i|\widetilde{0}\rangle_j|\widetilde{1}\rangle_k+\\
&\beta_7|\widetilde{1}\rangle_i|\widetilde{1}\rangle_j|\widetilde{0}\rangle_k+
\beta_8|\widetilde{1}\rangle_i|\widetilde{1}\rangle_j|\widetilde{1}\rangle_k)
-\frac{|v\rangle}{\sqrt{2}}\otimes
(\beta_1|\widetilde{0}\rangle_i|\widetilde{0}\rangle_j|\widetilde{0}\rangle_k\\
&+\beta_2|\widetilde{0}\rangle_i|\widetilde{0}\rangle_j|\widetilde{1}\rangle_k+
\beta_3|\widetilde{0}\rangle_i|\widetilde{1}\rangle_j|\widetilde{0}\rangle_k+
\beta_4|\widetilde{0}\rangle_i|\widetilde{1}\rangle_j|\widetilde{1}\rangle_k+\\
&\beta_5|\widetilde{1}\rangle_i|\widetilde{0}\rangle_j|\widetilde{0}\rangle_k+
\beta_6|\widetilde{1}\rangle_i|\widetilde{0}\rangle_j|\widetilde{1}\rangle_k+
\beta_7|\widetilde{1}\rangle_i|\widetilde{1}\rangle_j|\widetilde{0}\rangle_k\\
&-\beta_8|\widetilde{1}\rangle_i|\widetilde{1}\rangle_j|\widetilde{1}\rangle_k).
\end{split}\label{eq3}
\end{equation}
The measurement is made on the output photon by detectors D$h$ and
D$v$ behind PBS$d$. As PBS transmits (reflects) $h$ ($v$)-polarized
photon, according to Eq. (\ref{eq3}), the three-logic-qubit
conditional phase gate $\widetilde{U}_{\text{CP}3}$ succeeds if D$v$
clicks. With the help of single-logic-qubit $\widetilde{H}$
operation, we can carry out a standard Toffoli gate in DFS by
$\widetilde{T}_{ \text{offoli}}^{ijk}=\widetilde{H}_k\otimes
\widetilde{U}_{\text{CP}3}\otimes\widetilde{H}_k$. It must be
mentioned that the optical length of the paths $a$, $b$, $c$ and $i$
must be equal to suppress the phase instability in the single-photon
interference.

Different operations can be distinguished from the corresponding
ports and detectors. The single-logic-qubit operation
$\widetilde{H}_i$ in the $i$th cavity is associated with a
single-photon pulse input from the port 0$i$ and output to the
detector D$i$ \cite {wei}, whereas the many-qubit gating is
related to the input from the port 0$i$ and the output to the
detector D$v$. Since different gates could coexist in our design and
work independently, our design is not only scalable but also compact.

It is evident that our scheme can be directly extended to be more
than three-node case. Meanwhile, it could implement the conditional
phase flip between two arbitrary nodes. For example, by switching
TR$i$ (including TR$i$I and TR$i$II) to the reflecting state, we may
have the single-photon pulse skipping over PBS$a$. When it arrives
at the STR$j$, the single-photon pulse can also be chosen entering
node$j$ or not according to Fig.~\ref{2}(c). In this case, we can
implement the communications between two arbitrary nodes instead of
only the neighboring nodes \cite{wei}.

The currently achieved technology of deterministic single-photon
source \cite{hij}, with 10,000 high-quality single photons generated
continuously per second, supports a fast implementation of our
scheme. About the logic-qubits, we may confine the atoms in optical
lattices embedded in an optical cavity which has already been
achieved experimentally \cite{sauer}. But current techniques have
not yet enabled the atoms individually confined in some particular
lattice sites. Alternatively, we may consider two charged atoms
confined by a trap potential and optically coupled by the cavity
mode. A single Calcium ion has been successfully trapped in such a
device \cite {blatt}. To achieve our scheme, however, we require the
above experiment to be extended to two ions.

For the cavity-assisted operation $U_{\text{CPF}}$, the numerical
simulations had been made in Ref.\cite{cpf}, which shows that, if
the duration $T$ for the photon pulse input in the cavity and the
cavity decay rate $\kappa$ satisfy $\kappa T\gg 1$, $U_{\text{CPF}}$
is insensitive to both the atom-cavity coupling strength and the
Lamb-Dicke localization. Specifically, if $\kappa T\sim100$ and the
atom-cavity coupling is several times stronger than the dissipative
rates of the system, the gate fidelity is almost unity
($F>99.5\%$)\cite{cpf}. Therefore, with the experimental numbers
$\kappa/2\pi\sim4$ MHz, $g/2\pi\sim30$ MHz, $\Gamma/2\pi\sim2.6$ MHz
\cite{J,time}, we may estimate the time for $U_{\text{CPF}}$ and
$\widetilde{H}$ to be about $3\sim5\mu s$ and $6\sim10\mu s$,
respectively, for $\kappa T\gg1$. The time of the
$\widetilde{U}_{\text{CP}N}$ operation depends on the number of the
logic qubits. For example, for the case of $N=$3, 4 and 5, the
gating time of $\widetilde{U}_{\text{CP}N}$ would be about
$\sim12\mu s$, $\sim16\mu s$ and $\sim20\mu s$, respectively.

The possible imperfection is mainly from the photon loss,
the phase instability, detector inefficiency, and other logic
errors beyond collective dephasing in real QC operations, such as
leakage errors and so on. Like the repeat-until-success scheme
\cite{repeat} which discards the photon loss events by photon
detection, our scheme could reach high fidelity by the photon
detection even in the case of photon loss. Besides, the phase
stability could be guaranteed by keeping the path lengths of the
photons stable at sub-wavelength levels. Moreover, the current dark
count rate of the single-photon detector is about 100 Hz, which
could reduce the efficiency of our scheme by a factor of $10^{-4}$.
But this is not an intrinsic drawback of our scheme itself. As for
the logic errors beyond collective dephasing, we may suppress them
by some elaborately designed pulse sequences, e.g., with `Bang-Bang'
control pulses on the encoded qubits and then amended by refocusing
on individual physical qubits \cite {lind1}, or with some specially
designed pulses \cite {lind2}. To suppress these unpredictable
errors, we have to mention two points below. First, we have supposed
in our model that the collective dephasing errors are dominant for
the atomic qubits in a cavity at very low temperature, which implies
that the quantum gating with DFS employed in our scheme should
behave better than others without using DFS. The second point is
that both `Bang-Bang' control and refocusing have been sophisticated
techniques. So all the errors would be strongly suppressed.

In summary, we have proposed a many-logic-qubit network of cavity
QED, and demonstrated a three-logic-qubit Toffoli gate in DFS, which
could either carry out conditional gates between two arbitrary logic
qubits, or be extended to many-logic-qubit conditional gates. We
argue that our design is compact, scalable, and collective dephasing
resisted, which is practical in quantum network of cavity QED.

This work is partly supported by NNSFC under Grant Nos. 10774163 and
60490280, and partly by NFRP under grant No. 2006CB921203.

\end{document}